\documentclass[10pt,a4paper,twocolumn,english,american,aps,prl,twocolumn,showpacs,floatfix,superscriptaddress,10pt]{revtex4-1}
\usepackage[T1]{fontenc}
\usepackage[latin9]{inputenc}
\usepackage{babel}
\usepackage{array}
\usepackage{float}
\usepackage{multirow}
\usepackage{amssymb}
\usepackage{graphicx}
\PassOptionsToPackage{normalem}{ulem}
\usepackage[unicode=true,
 bookmarks=true,bookmarksnumbered=true,bookmarksopen=true,bookmarksopenlevel=2,
 breaklinks=false,pdfborder={0 0 0},backref=false,colorlinks=true]
 {hyperref}
\hypersetup{pdfauthor={Piotr Laczkowski },pdfsubject={Spintronics}}
\usepackage{breakurl}



\begin{document}

\title{Spin dependent transport characterization in metallic lateral spin
valves using 1D and 3D modeling}

%___________________________________AUTHORS
\author{P. Laczkowski}
\selectlanguage{english}%
\affiliation{Universit\'{e} Grenoble Alpes, CEA, CNRS, INP-G, INAC, F-38054 Grenoble, France}
\affiliation{Unit\'{e} Mixte de Physique CNRS/Thales, university Paris-Sud and Universit\'{e} Paris-Saclay, 91767 Palaiseau, France}

\author{M. Cosset-Cheneau}
\affiliation{Universit\'{e} Grenoble Alpes, CEA, CNRS, INP-G, INAC, F-38054 Grenoble, France}
\affiliation{D\'{e}partement de Physique, Ecole Normale Sup\'{e}rieure de Lyon, F-69342 Lyon France}

\author{W. Savero-Torres }
\affiliation{Universit\'{e} Grenoble Alpes, CEA, CNRS, INP-G, INAC, F-38054 Grenoble, France}

\author{V.T. Pham}
\affiliation{Universit\'{e} Grenoble Alpes, CEA, CNRS, INP-G, INAC, F-38054 Grenoble, France}

\author{H. Jaffr\`es }
\affiliation{Unit\'{e} Mixte de Physique CNRS/Thales, university Paris-Sud and Universit\'{e} Paris-Saclay, 91767 Palaiseau, France}

\author{N. Reyren }
\affiliation{Unit\'{e} Mixte de Physique CNRS/Thales, university Paris-Sud and Universit\'{e} Paris-Saclay, 91767 Palaiseau, France}

\author{J.-C. Rojas-S\`anchez }
\affiliation{Unit\'{e} Mixte de Physique CNRS/Thales, university Paris-Sud and Universit\'{e} Paris-Saclay, 91767 Palaiseau, France}

\author{A. Marty }
\affiliation{Universit\'{e} Grenoble Alpes, CEA, CNRS, INP-G, INAC, F-38054 Grenoble, France}

\author{L. Vila}
\affiliation{Universit\'{e} Grenoble Alpes, CEA, CNRS, INP-G, INAC, F-38054 Grenoble, France}

\author{J.-M. George }
\affiliation{Unit\'{e} Mixte de Physique CNRS/Thales, university Paris-Sud and Universit\'{e} Paris-Saclay, 91767 Palaiseau, France}

\author{J.-P. Attan\'{e} }
\affiliation{Universit\'{e} Grenoble Alpes, CEA, CNRS, INP-G, INAC, F-38054 Grenoble, France}

%___________________________________________Beginning of the paper:

\date{\today}
\selectlanguage{american}%
\begin{abstract}
We present the analysis of the spin signals obtained in NiFe based metallic lateral
spin valves. We exploit the spin dependent diffusive equations in
both the conventional 1D analytic modeling as well as in 3D Finite
Element Method simulations. Both approaches are used for extracting the spin diffusion length $l_{sf}^{N}$ and the effective spin
polarization $P_{eff}$ in Py/Al, Py/Cu and Py/Au based lateral nano-structures
at both $300\,K$ and $77\,K$. Both the analytic modeling and 3D Finite Element Method simulations give consistent results. Combination of both models provides a powerful tool for reliable spin transport characterization in all metallic spin valves and gives an insight into the spin/charge current and spin accumulations 3D distributions in these devices. We provide the necessary ingredients to develop the 3D finite element modeling of diffusive spin transport.

\end{abstract}
\maketitle

\section{Introduction}

Lateral Spin Valves (LSVs) are promising candidates for future spintronics
applications; to separate spin and charge current and to test the first two building blocks of a spin FET device \citep{DattaDas} : the spin injector and the spin detector. The ability to fabricate lateral devices allows, as well, to gain in design flexibility, of both ferromagnetic and non magnetic elements \citep{Zhand2017}. When characterizing spin dependent transport
in LSVs, many effects need to be taken into consideration, such as the quality of the interface, the interfacial resistance, the surface of contact, the spin flip at surfaces and interfaces, to name a few \citep{Mihajlovic2010a,Yakata_Kimura_milin_conditions_2010,Erekhinsky2010,Idzuchi_APL_2012,kimura2008tem}.
Moreover, one needs to take into account the possible
deviation of charge current path related to either the geometry \citep{Harmle2005}
or the difference in resistivities of the used materials. 3D modeling was proven to be essential for a proper analysis of the Spin Hall Effects \citep{Niimi_PRL_CuBi_2012}, and therefore its development is of a great importance. So far, 1D modeling has been mainly used for the quantitative estimation of the spin dependent transport in lateral spin valves, except in the work of Harmle \textit{et al} \cite{Harmle2005} where the authors used 3D network of resistors to analyze the spin signal of Py/Cu LSV's having wide Cu channels. In this paper we present the 3D modeling of spin dependent transport in the lateral metallic nano-structures based on the Finite Element Method simulations, and compare it with the standard 1D analytical modeling. Our study validate the 1D approach for a fast parameter extraction in metallic and interface transparent LSVs, and provides the necessary ingredients to develop such modeling using open access softwares. This allows us to report consistent values of the spin diffusion length of the non magnetic materials  and the effective spin polarization of Py in Py/Al, Py/Cu and Py/Au based lateral spin valves at both $300\,K$ and $77\,K$.

\section{Devices preparation and experimental results}

In order to characterize the spin dependent transport by extracting
the spin diffusion length $l_{sf}^{N}$ and the effective spin polarization
$P_{eff}$, Lateral Spin Valves (LSV) with different separation $L$
(from center to center) of the ferro-magnetic electrodes were fabricated
using both the multi-angle method \citep{Laczkowski2011} (Al and
Au based samples) or the multi-level method (Cu based samples). The
multi-level method consists in two sets of processes: lithography,
deposit and lift-off of the ferro-magnetic (F) material, followed
by the same steps for the non-magnetic (N) material. Importantly the multi-level
nano-fabrication process requires the cleaning of the
F/N interface by ion-milling before the deposition of the non-magnetic
channel. In contrast, in the case of the multi-angle evaporation technique
\citep{Valenzuela_Nature_2006,Yang2008a,Mihajlovic2010}, the sample
is kept in vacuum between the F and N wire depositions and hence for
the $F/N$ interface fabrication. This ensures good contacts quality,
without the need of interface cleaning between the deposition of the
ferromagnet and of the non-magnetic channel.

\begin{figure}
\begin{centering}
\includegraphics[width=7.7cm]{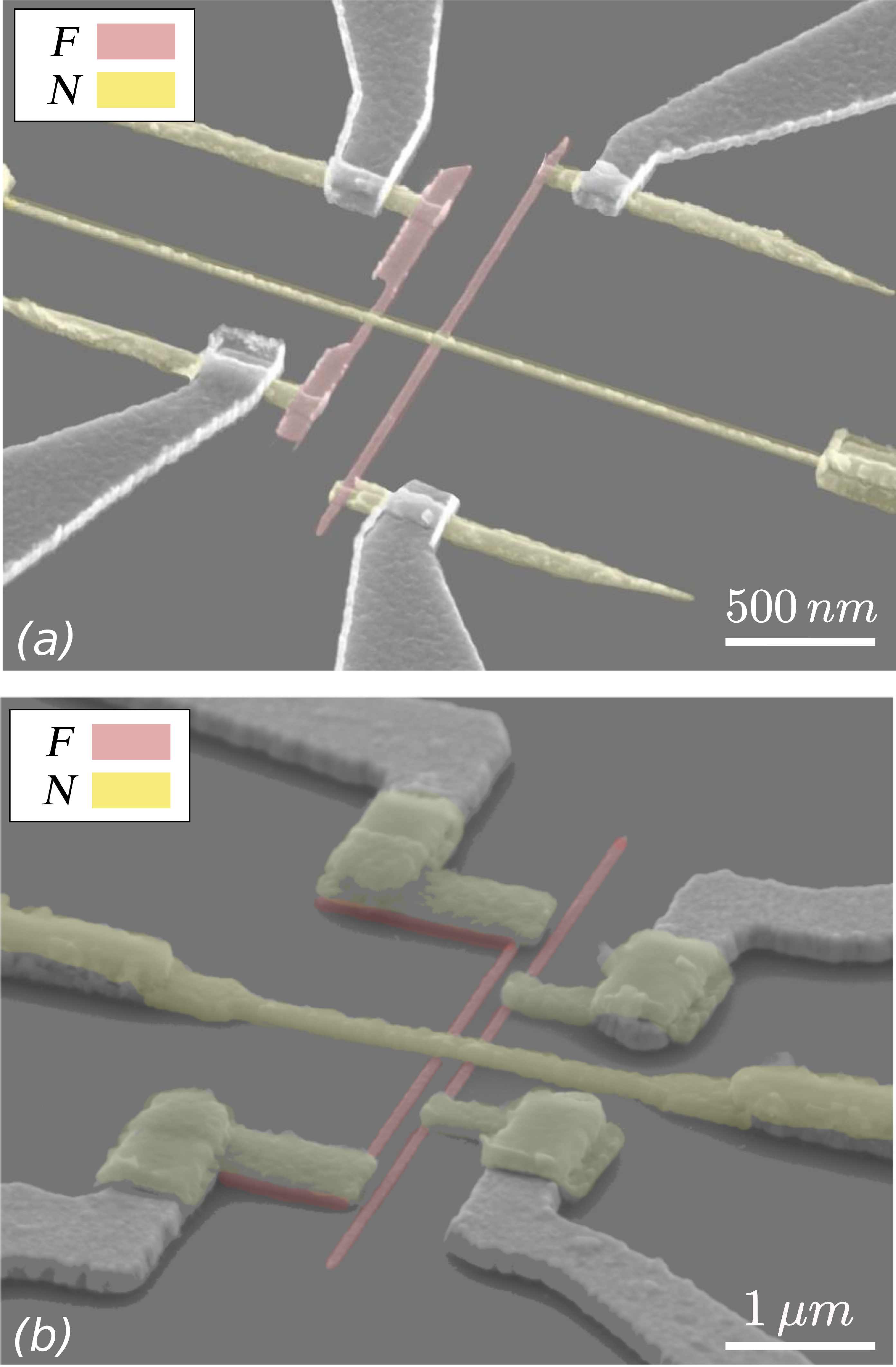}
\par\end{centering}

\caption{\textit{\label{fig:SEM-images}SEM image of the Lateral Spin Valve
device fabricated using (a) multi-angle evaporation technique, and (b)
multi-level technique. The ferro-magnetic electrodes and the non-magnetic
channel have been colored using the red and yellow colors respectively.}}
\end{figure}

First the $15\,nm$ thick and $50\,nm$ wide Permalloy stripes are
deposited on the silicone substrate, followed by the deposition of
the $50\,nm$ wide non-magnetic channel. In the case of the multi-angle
nano-fabrication method the thickness of N is $60\,nm$, and in the
case of multi-level method this thickness is increased to $80\,nm$.
Microscopic $Ti(5\,nm)/Au(100\,nm)$ contact electrodes are used to connect the active part of the devices. Figure \ref{fig:SEM-images} represents the Scanning Electron Microscope (SEM) images of typical nano-devices fabricated using (a)
the multi-angle and (b) the multi-level methods.

 For both nano-fabrication methods, in order to distinguish the switching
fields of the ferromagnetic electrodes (injector and detector), we added a nucleation
pad to one of the ferromagnets. This eases the nucleation process
of the electrode, allowing the switching of its magnetization direction at
a lower magnetic field \citep{Laczkowski2011}.

The spin signal amplitude has been measured as a function of the distance
$L$, varying from 100 nm to 1 $\mu$ m for samples with $Al$, $Cu$ and $Au$ non-magnetic channels, at both $300\,K$ and $77\,K$. For these three types of LSVs,
a standard lock-in amplifier measurement technique has been used with
$79\,Hz$ and $100\,\mu A$ ac current to measure the in-phase component
of the voltage output with a magnetic field oriented along the ferromagnetic
wires. The charge current injection and the voltage detection have been performed on the same side of the non-magnetic channel using the non-local technique depicted in fig. \ref{fig:Experimental_data}a. The values of the spin signal amplitudes, $\Delta R_{s}$ being the change of measured voltage divided by the injected current, are reported in Figure 2. This allows us to vary both the distance and the spin signals by at least one order of magnitude. We measured non-local spin signal ranging from $0.3$m$\Omega$ for Py/Au to $24$m$\Omega$ for Py/Cu based devices (fig. \ref{fig:Experimental_data}).

\section{Extraction of characteristic transport parameters from 1D model}

The spin diffusion length ($l_{sf}^{N}$) and the effective permalloy stripes
polarization ($P_{eff}$) are obtained by studying the gap dependence
of the spin signal. Assuming transparent interfaces between the non-magnetic channel and the ferromagnetic electrodes, the spin signal $\Delta R_s$ is expressed by a 1D spin diffusion model as \cite{Marty_2018}:

\begin{equation}
\Delta R_{s}=\frac{4 R_N \left(P_{eff}R_F \right)^2}{\left(R_N+2R_F \right)^2 e^{L/l_{sf}^N}-R_N^2 e^{-L/l_{sf}^N}}
\end{equation}

Here $R_{N(F)}=\rho_{N(F)}l_{sf}^{N(F)}/A_{N(F)}(1-P_{F}^{2})$ stands for the spin resistances, where: $A_{N}=w_{N}\times t_{N}$, $\rho_{N(F)}$, $l_{sf}^{N(F)}$, $t_{N}$,
$w_{N(F)}$, $A_{N(F)}$ are the resistivity, the spin diffusion length,
the thickness, the width and the cross sectional area. The F and N
subscripts  corresponds to the ferro-magnetic and non-magnetic material
respectively. Note that for the devices made by the multilevel method $A_F=w_F\times w_N$, while for those made by the multi-angle method, $A_F=(w_F+2t_F)\times w_N$ to take into account the conductivity of the sides of the electrodes, with $w_F$, $t_F$ the width and thickness of the electrode.

\begin{figure}
\begin{centering}
\includegraphics[width=7.4cm]{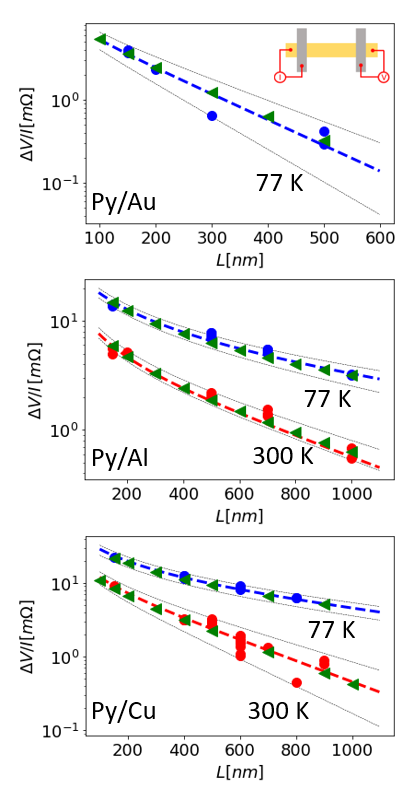}
\par\end{centering}

\caption{\textit{\label{fig:Experimental_data} Experimental data-points as
a function of distance $L$ for $77\,K$ (blue dots), altogether with
the fit results using the 1D model described by the equation (1)
(dashed curves) and the FEM simulations results (green triangles)
for (a) $Py/Au$ (b) $Py/Al$ and (c) $Py/Cu$ based nano-structures. The dark dashed lines corresponds to the spin signals obtained with the uncertainties given in table 1 for the corresponding material combinations. The insert in (a) corresponds to the non-local measurement configuration, the non-magnetic wire in yellow and the ferromagnetic electrodes in grey.}}
\end{figure}

Figure \ref{fig:Experimental_data} represents the experimental data-points
(dots) of the spin signal amplitude as a function of distance $L$
and fits using 1D (red and blue dashed lines) and 3D (green triangles) models for
(a) Py/Au, (b) Py/Cu and (c) Py/Al nano-structures. The data-points
recorded at $300\,K$ and $77\,K$ are represented by red and blue
colors respectively. In the case of $Py/Au$ based samples, only low-temperature and L $\leq$ 600 nm measurements were possible since the amplitude of the spin signal at $300\,K$ and large L was too small to be reliably detected. The estimated resistivities 
for Al, Cu and Au at room temperature are 
$30\,\Omega.nm$, $35\,\Omega.nm$, $35\,\Omega.nm$ and
at $77\,K$ are $15\,\Omega.nm$, $25\,\Omega.nm$,
$25-30\,\Omega.nm$ respectively. The resistivity and spin diffusion length of Py are $300\pm30 \,\Omega.nm$ and $5.2\pm2\, nm$ at room temperature, and at $10\, K$ are $220\pm12 \,\Omega.nm$ and $5.8\pm2\, nm$ \cite{Zahnd_2018}.

The experimental data-points have been fitted with two free parameters:
$P_{eff}$ and $l_{sf}^{N}$. $P_{eff}$ stands for the effective
spin polarization, it is basically a reduced bulk polarization as
it includes the depolarization by spin flip events at the interface:
the spin memory loss. $P_{eff}$ is then smaller than the obtained
bulk polarization. The spin memory loss can have various origins: spin
precession due to the magnetic stray field at rough interface, the
inter-diffusion between F and N materials or paramagnetic impurities \cite{SML}.

Summary of the fits results for the above mentioned samples-sets using
1D model are presented in table \ref{tab:FIT_sinh}. The longest
spin diffusion length $l_{sf}^{N}$ has been extracted for Py/Al then
for Py/Cu, and the smallest $l_{sf}^{N}$ was extracted for Py/Au
based structures. The extracted parameters for Py, Al and Cu are in good agreement with what can be found in literature for similar nano-structures\citep{Kimura2007a,Kimura2008,Wakamura_Otani_APEX_2011,Laczkowski_PRB_2012,Villamor2013}, and we find a quite long spin diffusion length ($\approx$ 140 nm) for Au at low temperature.

Regarding the effective spin polarization $P_{eff}$, the highest value at 77 K was extracted for Py/Cu and Py/Al and the smallest one for Py/Au based devices. This means that in the case of Py/Cu and Py/Al devices, the spin injection into the non-magnetic material is more efficient than in Py/Au at 77 K. It also appears that the spin injection efficiency is similar for Py/Al and Py/Cu at 300 K and 77 K.

\begin{table}[H]
\begin{centering}
\begin{tabular}{ccccc}
\hline 
$material$ & $P_{eff}$ & $l_{sf}^{N}\,[nm]$ & $\rho[\Omega \, nm]$ & $T\,[K]$\tabularnewline
\hline 
\hline 
\multirow{2}{*}{$Py/Al$} & $0.22\pm0.01$ & $450\pm90$ & $30$ &$300$\tabularnewline
 & $0.32\pm0.01$ & $1100\pm300$ & $15$ & $77$\tabularnewline
\multirow{2}{*}{$Py/Cu$} & $0.22\pm0.01$ & $300\pm60$ & $35$ &$300$\tabularnewline
 & $0.31\pm0.03$ & $900\pm80$ & $25$ & $77$\tabularnewline
$Py/Au$ & $0.26\pm0.01$ & $140\pm30$ & $27\pm3$ &  $77$\tabularnewline
\hline 
\end{tabular}
\par\end{centering}

\caption{\label{tab:FIT_sinh}\textit{Representation of the fit results (for
a given temperature) obtained by using equation (1)
for $Py/Al$, $Py/Cu$ and $Py/Au$ LSV sample's data-sets.}}
\end{table}

\section{Finite Elements Method simulations}

The FEM simulations have been performed using: GMSH \citep{gmsh},
a three-dimensional finite element mesh generator, and GetDP the finite
element method (FEM) solver \citep{getdp_base,getdp_ieee_1999}. A
collinear approach has been used for simplicity, which is equivalent
to the case of two opposite magnetization orientations along a chosen
axis. It allows us to access the situation of parallel and anti-parallel
alignment of the ferromagnetic electrodes. In this formulation we
choose a current density $j$ to represent a flow of spins, otherwise
it would represent the electric current density, and the sign of
following equations would have to be changed.

\paragraph{Formulation}

The FEM calculations are based upon the diffusive transport equations,
where the currents of carriers with up and down spins have been derived
from the electro-chemical potentials $\mu_{\uparrow(\downarrow)}$,
with different conductivities $\sigma_{\uparrow}$ and $\sigma_{\downarrow}$.
In this image the current densities can be expressed in the following
form:

\begin{equation}
\begin{array}{c}
\overrightarrow{j_{\uparrow}}=\sigma_{\uparrow}\overrightarrow{\nabla}\mu_{\uparrow}=\sigma\frac{1+P}{2}\overrightarrow{\nabla}\mu_{\uparrow}\\
\\
\overrightarrow{j_{\downarrow}}=\sigma_{\downarrow}\overrightarrow{\nabla}\mu_{\downarrow}=\sigma\frac{1-P}{2}\overrightarrow{\nabla}\mu_{\downarrow}
\end{array}\label{eq:j_up_and_down}
\end{equation}

where $\sigma$ is defined as $\sigma=\sigma_{\uparrow}+\sigma_{\downarrow}$,
representing the total conductivity, and $P$ is the current polarization.
The charge current conservation imposes that

\begin{equation}
div(\overrightarrow{j_{\uparrow}}+\overrightarrow{j_{\downarrow}})=0\label{eq:div_j=00003D0}
\end{equation}

what can be further merged with a spin relaxation proportional to
the spin accumulation in the following form:

\begin{equation}
div\left(\overrightarrow{j_{\uparrow}}\right)=-div\left(\overrightarrow{j_{\downarrow}}\right)=\alpha\left(\mu_{\uparrow}-\mu_{\downarrow}\right)=\frac{1-P^{2}}{4\rho l_{sf}^{2}}\left(\mu_{\uparrow}-\mu_{\downarrow}\right)\label{eq:form1}
\end{equation}

where $\rho=1/\sigma=1/(\sigma_{\uparrow}+\sigma_{\downarrow})$ is
the global resistivity and $l_{sf}$ is the spin diffusion length.

By combining equations \ref{eq:j_up_and_down} with \ref{eq:form1},
one recovers the well known diffusion equation \citep{ValetFert1993}:

\begin{equation}
\triangle(\mu_{\uparrow}-\mu_{\downarrow})=\frac{\mu_{\uparrow}-\mu_{\downarrow}}{l_{sf}^{2}}
\end{equation}

For transparent interfaces, the continuity conditions on the interfaces
are imposed (continuity of the electro-chemical potential), together
with the normal spin current densities continuity at the interfaces.
The material connecting the terminals is assumed to be long enough, to have vanishing spin accumulation on the terminal side. We typically choose this length to be at least three times $l_{sf}$ and/or three times its width. The later allows the charge current to be homogeneously spread over the section of the wire far away from the interfaces. One thus assumes the same polarization on terminal faces, than in the bulk material:

\begin{equation}
\begin{array}{c}
j_{\uparrow surf}=\frac{1+P}{2}\frac{I}{A}\\
\\
j_{\downarrow surf}=\frac{1-P}{2}\frac{I}{A}
\end{array}
\end{equation}

The spin dependent transport can then be defined in terms of the charge
($j_{c}$) and spin ($j_{s}$) currents by the mean and the difference
of the electro-chemical potentials:

\begin{equation}
\begin{array}{c}
\overrightarrow{j_{c}}=\overrightarrow{j_{\uparrow}}+\overrightarrow{j_{\downarrow}}=\sigma\overrightarrow{\nabla}\left(\frac{\mu_{\uparrow}+\mu_{\downarrow}}{2}\right)+P\sigma\overrightarrow{\nabla}\left(\frac{\mu_{\uparrow}-\mu_{\downarrow}}{2}\right)\\
\\
\overrightarrow{j_{s}}=\overrightarrow{j_{\uparrow}}-\overrightarrow{j_{\downarrow}}=\sigma\overrightarrow{\nabla}\left(\frac{\mu_{\uparrow}-\mu_{\downarrow}}{2}\right)+P\sigma\overrightarrow{\nabla}\left(\frac{\mu_{\uparrow}+\mu_{\downarrow}}{2}\right)
\end{array}
\end{equation}

\paragraph{Mesh density:}

the evaluation of the optimum mesh density is a very important aspect of the FEM simulations. The results of the simulations can be significantly changed when a too low tetrahedron number is used in order to mesh a given geometry. This become important especially in the case of geometries including right angles and edges.

\begin{figure}
\begin{centering}
\includegraphics[width=8cm]{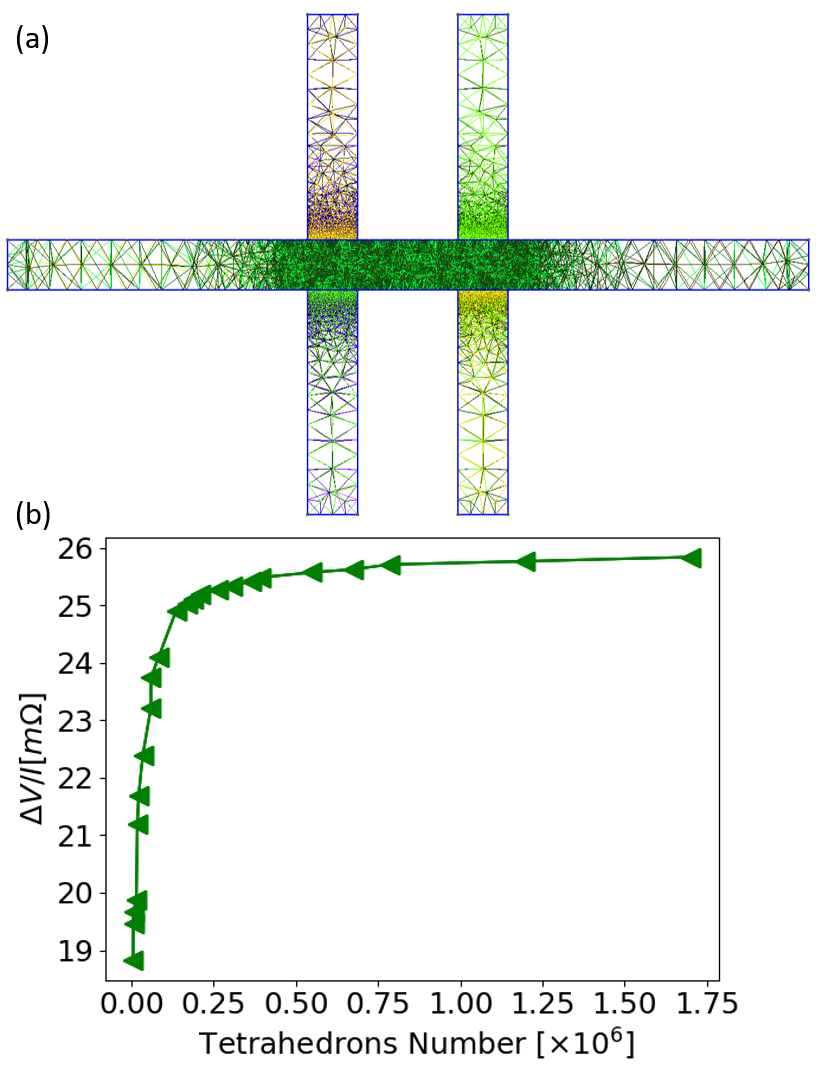}
\par\end{centering}

\caption{\textit{\label{fig:mesh}(a) Mesh density used for the simulations in the lateral spin valves. The mesh density is larger in the active part (center) of the device where the right angle points and edges are concentrated. (b) Variation of the signal as a function of mesh density (expressed
by the total number of tetrahedrons). The output signal corresponds to the spin signal obtained with the 3D model using the Py/Al parameters with a geometry representing a multi-level fabrication method. The spin signal depends on the mesh density for low tetrahedrons number, and stabilizes when the meshing is sufficiently refined. For representation purposes, the length of the metallic wires have been taken shorter than the ones used in the calculations.}}
\end{figure}

In our simulations, the distance between vertices at the proximity of edges and right angles has been chosen to be $1.2$ nm, which corresponds to a number of $0.5 \cdot 10^6$ tetrahedrons for a non-magnetic channel of section $50$ nm $\times \,80$ nm with a distance of $150$ nm between the electrodes. This ensures a low variation of the output signal as shown in fig. \ref{fig:mesh}b when increasing the mesh density while keeping a reasonable computation time. The refined meshing area extends to a distance of $120$ nm from the edges and right angles of the structure, and the distance between the vertices changes continuously to attain a value of $30$ nm outside the active part of the structure (fig \ref{fig:mesh}a). We checked that the output signal does not depend on the distance between the vertices outside the active part of the structure.

\paragraph{Fitting using 3D model:}

\begin{figure}
\begin{centering}
\includegraphics[width=8cm]{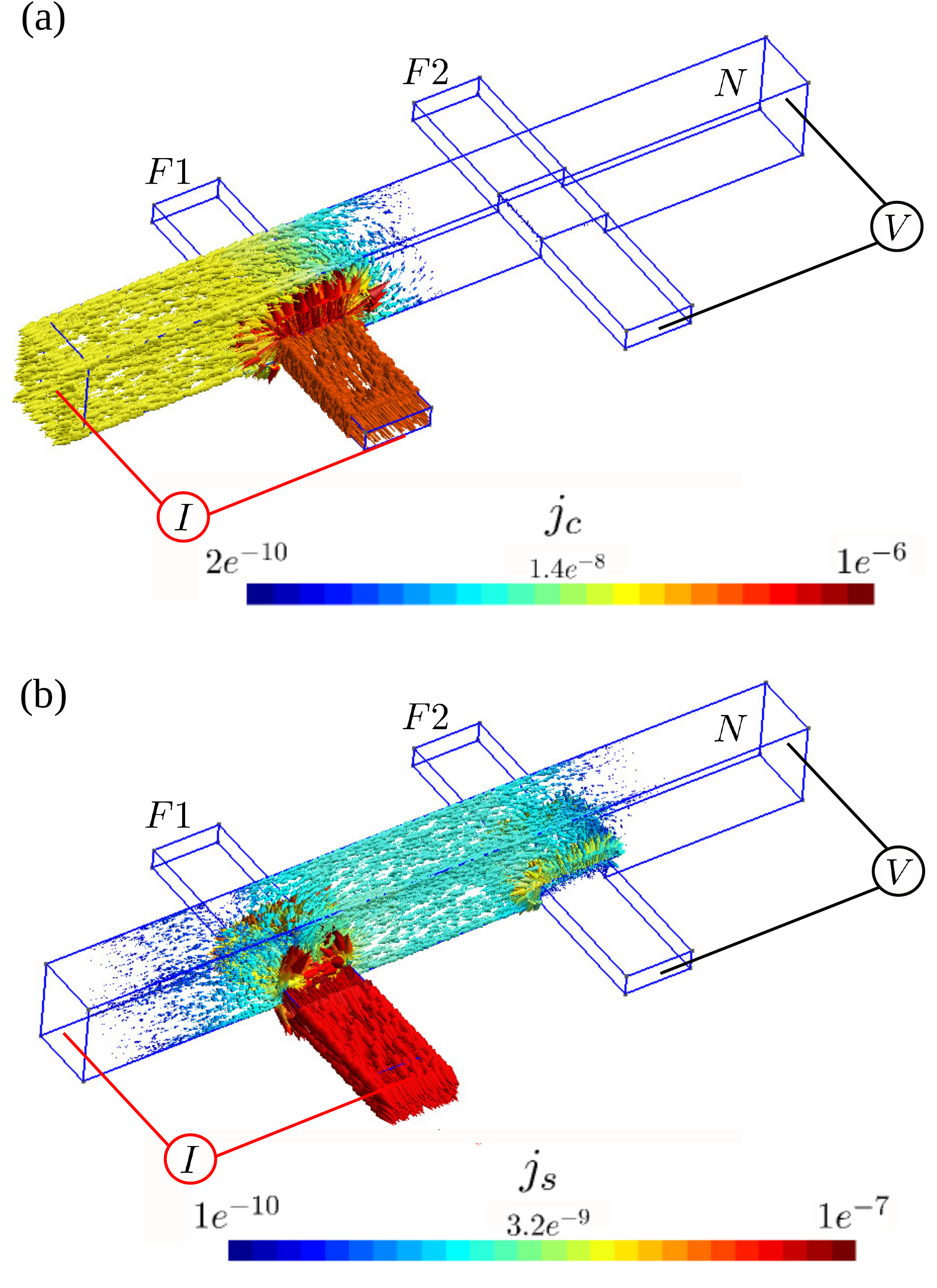}
\par\end{centering}

\caption{\textit{\label{fig:FEM1} FEM simulations results for a given geometry
of the nano-structure with (a) charge current $j_{q}$ injected at
the bottom of the ferro-magnetic electrode F1 (bottom part) and drained
out through the left side of a non-magnetic material, and (b) spin
current $j_{s}$, with the efficient absorption of F2, reflecting
the situation with $R_{F}<R_{N}$. Both $j_{c}$ and $j_{s}$ are
displayed in the logarithmic scale. }}
 
\end{figure}

In the 3D simulations, the distribution of the charge current, the
spin current and the spin current accumulation have been calculated
for two magnetic configurations (parallel P and anti-parallel
AP). The spin signal amplitude has been reproduced by taking the difference
of the output signals for the two states. In each case, the spin signal
is evaluated from the difference of electro-chemical potentials integrated
on the end surface of the voltage contacts (fig. \ref{fig:FEM1}).
The contact wires have to be long enough (several $l_{sf}$) to cancel
spin accumulation, so both up and down electro-chemical potentials
are equal to the pure electric potential. This corresponds to the
non-Local probe configuration setup detection, where the voltage is
probed between the right side of the non-magnetic channel and the
lower part of the second ferromagnetic wire. 

The charge current is injected in the left side of the device through
the first ferromagnetic wire and flows at the F1/N interface. It is then
drained out on the left side of N. This situation is represented in
figure \ref{fig:FEM1}(a), where the charge current $j_{q}$ is displayed
using colored arrows. Thus, the created spin accumulation diffuses
in the non-magnetic channel, creating the spin currents $j_{s}$. The
distribution of spin currents is represented in figure \ref{fig:FEM1}(b) for the case of the AP
magnetic state. 

\begin{figure}
\centering{}\includegraphics[width=8cm]{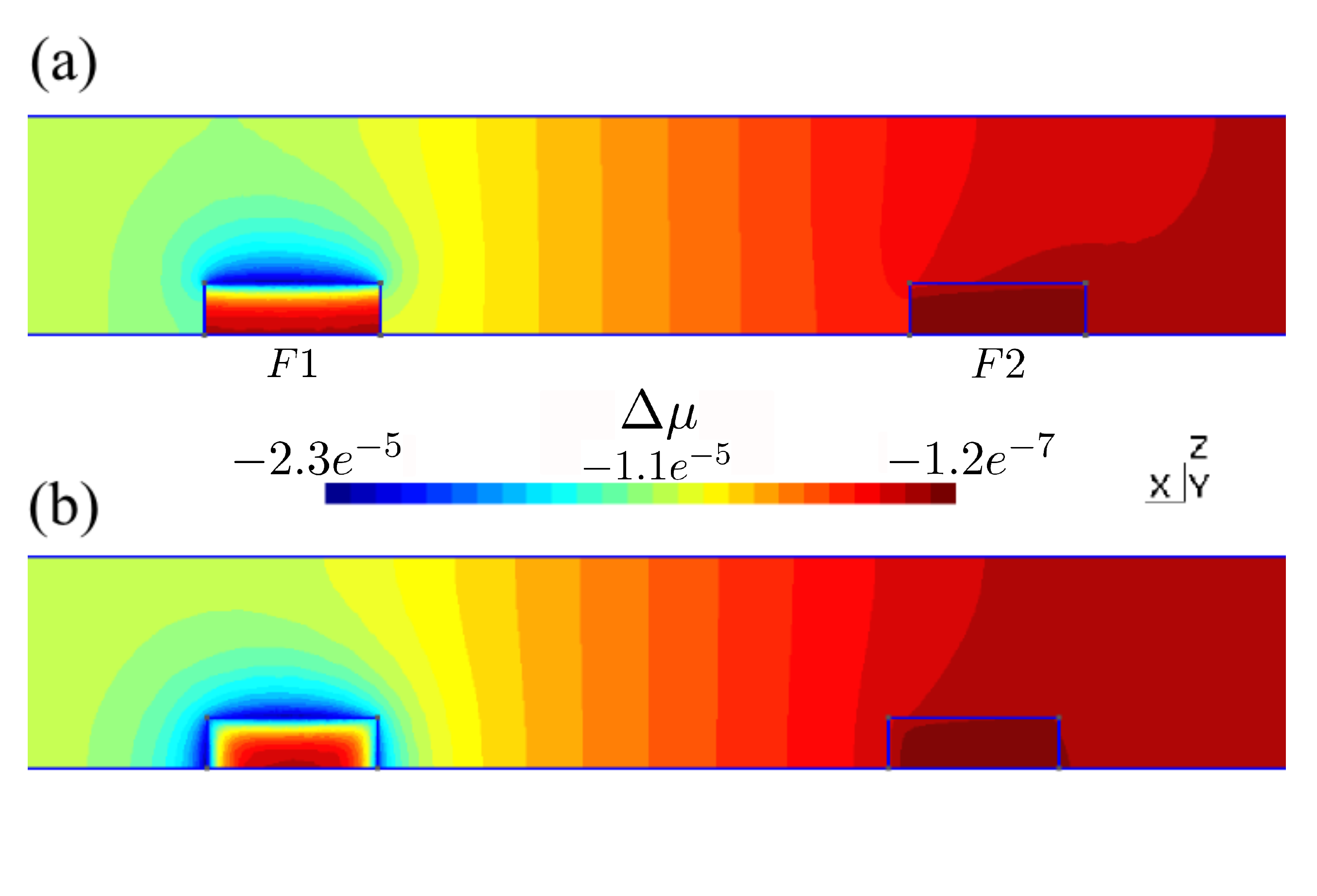}\caption{\textit{\label{fig:FEM2}FEM simulations results of the spin accumulation
distribution $\mu_{a}=\mu_{\uparrow}-\mu_{\downarrow}$ in a LSV on
the longitudinal cut at the center of N. The charge current is injected
on the left side of the ferro-magnetic material (F1) while the difference
of the spin accumulation (for a given magnetic state of the system
that can be parallel or anti-parallel) is measured between F2 and
the right side of N. All values are displayed by using the iso-surfaces
representation in the logarithmic scale.}}
\end{figure}

The resulting spin accumulation distribution $\mu_{a}=\mu_{\uparrow}-\mu_{\downarrow}$ for the AP magnetic state is represented in figure \ref{fig:FEM2}(a-b),
using the ISO-surface representation. Note that two cases have been
considered depending on either the experimental data-points were extracted
from the multi-angle or from the multi-level nano-fabrication method.
Figure \ref{fig:FEM2}(a) stands for the case where the charge current
is injected into N through the top surface of the ferro-magnetic electrode
F1 which represents the multi-level nano-fabrication method. Only the
top surface is cleaned before the deposition of the non-magnetic channel,
and the sides of F1 do not contribute to the current
injection. Figure \ref{fig:FEM2}(b) represents the case describing
the multi-angle nano-fabrication method where the active part of the
device is evaporated in a single step, without breaking a vacuum,
and therefore all surfaces of contact between F and N need to be taken
into account in the current injection process analysis.

Table \ref{tab:FEM-fit_results} summarizes 3D model fitting results from fig.  \ref{fig:Experimental_data} for Py/Al, Py/Cu and Py/Au nano-structures, assuming transparent interfaces. For all samples, the values for the effective polarization $P_{eff}$ and spin diffusion length $l_{sf}^N$ are only slightly larger than the ones found with the 1D model. This validate the 3D simulations and confirms the 1D approximation in our structures. 

\begin{table}[H]
\begin{centering}
\begin{tabular}{ccccc}
\hline 
$material$ & $P_{eff}$ & $l_{sf}^{N}\,[nm]$ & $\rho[\Omega \, nm]$ & $T\,[K]$\tabularnewline
\hline 
\hline 
\multirow{2}{*}{$Py/Al$} & $0.24\pm0.01$ & $496$ & $30$ &$300$\tabularnewline
 & $0.33\pm0.01$ & $1100$ & $15$ & $77$\tabularnewline
\multirow{2}{*}{$Py/Cu$} & $0.24\pm0.01$ & $300$ & $35$ & $300$\tabularnewline
 & $0.35\pm0.01$ & $900$ & $25$ & $77$\tabularnewline
$Py/Au$ & $0.27\pm0.01$ & $150$ & $30$ & $77$\tabularnewline
\hline 
\end{tabular}
\par\end{centering}

\caption{\label{tab:FEM-fit_results}\textit{The 3D models fitting results
of the $l_{sf}^{N}$ and $P_{eff}$ allowing to reproduce the experimental
data-points.}}
\end{table}

\section{Summary from two models}

Parameters estimated using both above described characterization methods
(1D and 3D) are summarized in table \ref{tab:summary_fits}. Both
the $P_{eff}$ and $l_{sf}^{N}$ values are in good agreement with
what can be found in literature for similar nano-structures \citep{Jedema2003,Kimura2005PRB,Costache_PRB_2006,Ji20071280,Kimura2008}.

The obtained values for the effective polarization and the spin diffusion lengths obtained by 1D and 3D analysis are in good agreement. The slightly larger values obtained with the 3D model for $P_{eff}$ may be due to the additional relaxation linked with the spatial extension of the injector. This effect is not taken into account by the 1D model that assumes a point contact between the ferromagnetic electrodes and the non-magnetic channel \cite{PiotrAuW2015}. This leads the 1D model to overestimate the spin-signal, hence leading to a lower value for $P_{eff}$. 

At room and low temperature the $l_{sf}^N$ values were found to be highest for Py/Al, then for Py/Cu and finally for Py/Au based devices. 
This means that at low temperatures the spin currents propagate to
longer distances in Al than in Cu followed by Au channels.
The effective polarization also depends on the materials combination. 
At low temperature, $P_{eff}$ is larger for Py/Cu and Py/Al based devices than in the Py/Au based ones. This indicates a better spin injection efficiency for the Py/Cu and Py/Al interfaces. 

%At room temperature $P_{eff}$ is larger for Py/Cu based devices than in the Py/Al based one. This indicates a better spin injection efficiency related to a lower interface resistance for the Py/Cu interface as measured in \cite{PhDPiotr}.

\begin{table}[H]
\begin{centering}
\begin{tabular}{ccccc}
\hline 
$material$ & $model$ & $P_{eff}$ & $l_{sf}^{N}\,[nm]$ & $T\,[K]$\tabularnewline
\hline 
\hline 
\multirow{4}{*}{$Py/Al$} & $1D$ & $0.22\pm0.01$ & $450\pm90$ & \multirow{2}{*}{$300$}\tabularnewline
 & $3D$ & $0.24\pm0.01$ & $496$ & \tabularnewline
 & $1D$ & $0.32\pm0.01$ & $1100\pm300$ & \multirow{2}{*}{$77$}\tabularnewline
 & $3D$ & $0.33\pm0.01$ & $1100$ & \tabularnewline
\hline 
\multirow{4}{*}{$Py/Cu$} & $1D$ & $0.22\pm0.01$ & $300\pm60$ & \multirow{2}{*}{$300$}\tabularnewline
 & $3D$ & $0.24\pm0.01$ & $300$ & \tabularnewline
 & $1D$ & $0.31\pm0.03$ & $900\pm80$ & \multirow{2}{*}{$77$}\tabularnewline
 & $3D$ & $0.35\pm0.01$ & $900$ & \tabularnewline
\hline 
\multirow{2}{*}{$Py/Au$} & $1D$ & $0.26\pm0.01$ & $140\pm30$ & \multirow{2}{*}{$77$}\tabularnewline
 & $3D$ & $0.27\pm0.01$ & $150$ & \tabularnewline
\hline 
\end{tabular}
\par\end{centering}

\caption{\label{tab:summary_fits}\textit{Summary of the fit results from 1D
and 3D models, for Py/Al, Py/Cu and Py/Au sample sets. }}
\end{table}
\section{Comparison between 1D and 3D modeling}
\begin{figure}
\centering{}\includegraphics[width=8cm]{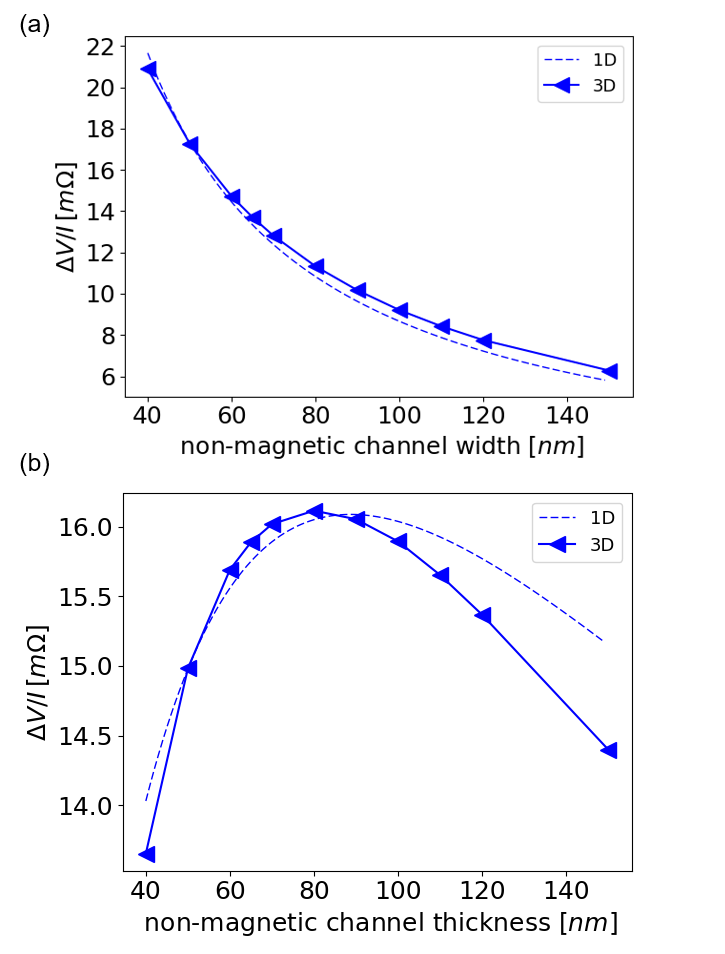}\caption{\textit{\label{fig:geo} Comparison of the predicted evolution of the spin signal when varying the width (a) and the thickness of the non-magnetic channel (b) for 1D and 3D modeling. The used parameters have been obtained from 1D and 3D modeling for Py/Cu at 77 K. The measurement contacts have been taken on the same side of the device for the 3D model.}}
\end{figure}

We have seen in the previous section that the FEM simulations in our devices give consistent results for the materials parameters when the devices are fabricated using the multi-level method. However, the 1D model assumes a uniform distribution of the spin current across the non-magnetic channel section. It has been observed \cite{Kimura_Otani2007} that a variation of the non-magnetic channel cross section can lead to a large variation of the predicted spin signal. In order to further explore the agreement of the 1D and 3D analysis, we compared the calculated spin signal obtained by the 1D and 3D model when varying the channels cross section. 
Increasing the width $w_N$ (fig. \ref{fig:geo}a) of the channel decreases the spin resistance of the ferromagnetic injector, resulting in a reduction of $\Delta R_s \propto 1/w_N$ according to equation (1) due to a lower spin injection efficiency \cite{Kimura_Otani2007}. On the other hand, an increase in the thickness $t_N$  of the non-magnetic channel leads to a reduction of the spin resistance mismatch, eventually leading to a better spin injection. Nevertheless, this provide more room for the spin accumulation to vanish without being detected, inducing the non-monotonic shape observed in fig. \ref{fig:geo}b. This large relaxation volume effect is not taken into account by the 1D model, which explains the discrepancy with the 3D modeling  for $t_N>w_N$.

We observed overall a very good agreement between the two models when varying the width of the non-magnetic channel (fig \ref{fig:geo}.a), and a good agreement with difference of less than 10\% between the 1D and 3D predictions when varying the thickness of the non-magnetic channel (fig \ref{fig:geo}.b).

This validate the use of the 1D model for LSVs fabricated using the multi-level method even for large non-magnetic channel width when its thickness is not too wide. However, for thick non-magnetic channel the 3D model should be used to obtain more accurate results. A precise evaluation of the spin signal change when changing the cross section of the non-magnetic channel will be a key element for the precise evaluation of the spin diffusion length when taking into accounts spin-flip effects at the non-magnetic channel surfaces \cite{Erekhinsky2010}.

\section{Conclusions}

In conclusion, we have analyzed the spin dependent transport parameters of Py based LSV's, having a spin channel made of Al, Cu and Au at T=77 K and 300 K. We compared the results obtained by the conventional 1D analytical modeling of the spin transport across transparent interface to the results obtained by a 3D resolution of the problem using Finite Element Modeling. The results of both analyzes are consistent, which validates the use of 1D modeling for a fast extraction of the material parameters in the studied cases. 
This allows providing quite robust material parameters $l_{sf}^{N}$ and $P_{eff}$, of LSVs with Py electrodes and Al, Cu and Au non-magnetic channels, and appears to be in quite good agreement with previous experiments found in literature. The development of FEM of spin transport appears as an essential tool for further use in devices with complex geometries where 1D modeling is a too strong assumption as observed in the case of thick non-magnetic channel. In this paper, we provide the key ingredients required for the development of such FEM methods for studying diffusive spin transport.

\begin{acknowledgments}
The devices were fabricated at RENATECH Grenoble. We acknowledge financial support from the Agence Nationale de la Recherche, ANR-16-CE24-0017 TOPRISE and the Laboratoire d'Excellence LANEF (ANR-10-LEBX-51-01).
\end{acknowledgments}

\end{document}